\documentclass[12pt,a4paper]{article}
\usepackage{acromake}
\let\vecc\mathbf                              
\def\vprod{\mathbin{\null_\wedge}}           
\def\Tr{\mathrm{Tr}}                         
\def\Re{\mathrm{Re}}                         
\def\Im{\mathrm{Im}}                         
\def\im{\mathrm{i}}                          
\def\d{\mathrm{d}}                           
\def\e{\mathrm{e}}                           
\def\eg{\emph{e.g.}}
\acromake{WW}{WW}{Weizs{\"a}cker-Williams}
\title{Charge-odd and single-spin effects in
\\
  two-pion production in $e\vec{p}$ collisions%
\footnote{The work of EAK and BGS was partially supported by RFBR grant
  ~99-02-17730 and HLP grant~2000-02.} }

\author{
Mikhail Galynsky$^{1}$
\ ,
{\'E}duard A. Kuraev$^{2}$
\ ,\\
Philip G. Ratcliffe$^{3,}$\thanks{E-mail: \texttt{pgr@fis.unico.it}} \
and
Binur G. Shaikhatdenov$^{2,}$\thanks{E-mail: \texttt{sbg@thsun1.jinr.ru}}
\\[12pt]
  \hskip-1.5em
  \parbox{\textwidth}
  {
\begin{list}{$^\arabic{enumi}$}
          {\small
           \topsep0pt
           \partopsep0pt
           \parsep0pt
           \itemsep0pt
           \usecounter{enumi}%
           \labelwidth1.1ex
           \def\makelabel##1{##1\hss}%
           \it
          }
  \item Stepanov Institute of Physics BAS, 220072, Minsk, Skorina ave.,
        70, Belarus
  \item Joint Inst.\ for Nuclear Research,
        141980, Dubna, Moscow region, Russia
  \item Dip.\ di Scienze CC.FF.MM., Univ.\ degli Studi dell'Insubria,
        via Lucini 3, 22100 Como, Italy
        and Ist.\ Naz.\ di Fisica Nucleare---sezione di Milano
  \end{list}}}
\date{}
\begin{document}
\maketitle
\begin{abstract}
We consider double-photon and bremsstrahlung mechanisms for the production of
two charged pions in high-energy electron (or proton) scattering off a
transversely polarised proton. Interference between the relevant amplitudes
generates a charge-odd contribution to the cross-section for the process. In
the kinematical configuration with a jet nearly collinear to the electron,
the spin-\emph{independent} part may be used to the determine phase
differences for pion-pion scattering in states with orbital momentum 0 or 2
and 1, while for the configuration with a jet nearly collinear to the proton,
the spin-\emph{dependent} part may be used to explain the experimental data
for single-spin correlations in the production of negatively charged pions.
We also discuss the backgrounds and estimate the accuracy of the results to be
better than 10\%. In addition, simplified formul{\ae} derived for specific
kinematics, with small total transverse pion momenta, are given.
\end{abstract}
\newpage
\section{Introduction}

Among the possible methods of measuring pion-pion scattering phases at low
energies~\cite{Dafne:1998x1} (\eg, $K_{e4}$ decays, pionium atoms,
pion-to-two-pion transitions in the scattering of pions by protons---the
Chew-Low process), high-energy processes in which a jet consisting of pions
with relative small invariant mass is created have not, up to now, been
considered. The theoretical possibility was, however, considered some time ago
by Serbo and Chernyak~\cite{Serbo:1974x1} for a special kinematical region of
the produced pions.

In this paper we calculate the charge-odd contribution to the cross-section in
charged-pion pair production for the kinematics of a jet moving close to the
direction of one of the initial particles. We consider the general case,
restricted only by the requirement that the sum of the momenta transverse to
the beam axis be close to zero. This region corresponds to the case where one
of the scattered particles moves close to its initial direction and escapes
detection while the components of the jet moving in the opposite direction
have measurable scattering angles and are assumed to be detected.

The bremsstrahlung mechanism of pion pair creation includes the conversion of
a time-like photon into a pion pair via an intermediate $\rho$-meson state.
The Breit-Wigner resonance form of the relevant pion form factor provides an
imaginary part, which can give rise to single-spin correlation effects in the
differential cross-section. These last may also arise as interference effects
between the Born and one-loop Feynman diagrams for single-pion
production~\cite{Ahmedov:1999ne} beyond the resonance region, due to the
presence of intermediate nucleon resonance states~\cite{Arbuzov:1994x1}. The
charge-odd contribution to the cross-section has a clear signal: the
invariant mass of two pions equals the $\rho$-meson mass and may thus be
separated from the even part of the cross-section in a two-meson exclusive
set-up.

The idea of measuring the distributions in the fragmentation region of one of
the colliding beams was first considered in papers of the
seventies~\cite{Kuraev:1976x2}. In particular, the odd part of the
cross-section for the production of a muon pair in a jet moving along the
initial-electron direction was obtained there. The energy distributions in a
jet moving along, for example, the initial-electron direction were obtained in
the form of a Dalitz-plot for jets consisting of two electrons and one
positron and for those of one electron and two photons. For instance, the
charge-odd contribution to the spectrum for muon-pair production takes the
form
\begin{eqnarray}
  \frac{\d^2 \sigma^{e^+e^- \to e^+e^-\mu^+\mu^-}_\mathrm{odd}}
       {\d{x}_+ \, \d{x}_-}
  &=&
  \frac{4\alpha^4}{3\pi m_\mu^2} \,
  \ln\left(\frac{s}{m_em_\mu}\right)
  \frac{x_+x_-(x_+-x_-)(2-x_+-x_-)}{(x_++x_-)^4}
\nonumber\\
  &=&
  0.58 \, \mathrm{nb} \; \cdot \; f_4(x_+,x_-),
\label{f4}
\end{eqnarray}
with
\begin{equation}
  x_{\pm} = \frac{E_{\pm}+p^z_{\pm}}{2E},
  \qquad
  s = 4E^2,
\end{equation}
where $E$ is the electron beam energy in the centre-of-mass reference frame;
$x_{\pm}$ are the muon energy fractions; $E_{\pm}$, $p^z_{\pm}$ the energies and
$z$-components of the muon momenta and $m_e$ and $m_\mu$ are the electron and
muon masses respectively.

Bearing in mind that the even part of the cross-section has a similar order
of magnitude~\cite{Baier:1981x1}, one sees that the asymmetry effects are of
order unity. We first consider the jet-2 kinematics, specified by
interference of the amplitudes associated with the diagrams depicted in
Figs.~\ref{fig:diags}a (double-photon mechanism) and~\ref{fig:diags}b
(bremsstrahlung along $p_2$). The case of the jet-1 kinematics, characterised
by interference of the amplitudes graphically represented in
Figs.~\ref{fig:diags}a and~\ref{fig:diags}c, is then elaborated in some
detail. Then, we derive reduced formul\ae, for a back-to-back kinematics with
pions moving almost in opposite directions, of the expressions obtained
earlier. In conclusion we apply our results to a particular experimental
setup (HERA).

\section{Jet-2 kinematics}

In this paper we deal with the charged-pion pair production channel in
high-energy collisions between unpolarised electrons and transversely
polarised protons. We first examine the kinematical region for the creation
of a jet moving close to initial proton direction:
\begin{eqnarray}
  e(p_1) + p(p_2,a)
  \to
  e(p_1') + p(p_2') + \pi^+(q_2) + \pi^-(q_1),
\end{eqnarray}
\begin{eqnarray*}
  p_1^2 = p_1^{\prime2} = m_e^2,
  \quad
  p_2^2 = p_2^{\prime2} = M^2,
  \quad
  q_{1,2}^2 = m^2,
  \quad
  p_2 {\cdot} a = 0,
\nonumber
\end{eqnarray*}
where $m_e$, $M$ and $m$ are the electron, proton and pion masses
respectively and $a$ is the proton polarisation 4-vector.
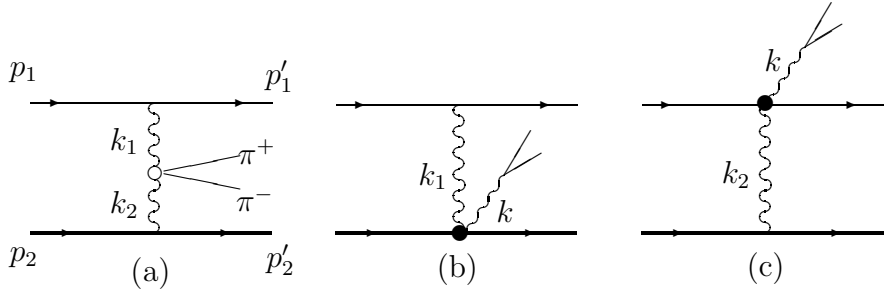
\begin{figure}
\unitlength=2.00pt
\special{em:linewidth 0.4pt}
\begin{picture}(174.31,55.88)
  \put(36.29,13.30){\oval(2.00,2.00)[r]}
  \put(36.29,15.30){\oval(2.00,2.00)[l]}
  \put(36.29,17.30){\oval(2.00,2.00)[r]}
  \put(36.29,19.30){\oval(2.00,2.00)[l]}
  \put(36.29,25.87){\oval(2.00,2.00)[r]}
  \put(36.29,27.87){\oval(2.00,2.00)[l]}
  \put(36.29,29.87){\oval(2.00,2.00)[r]}
  \put(36.29,31.87){\oval(2.00,2.00)[l]}
  \put(36.29,33.87){\oval(2.00,2.00)[r]}
  \put(36.29,35.87){\oval(2.00,2.00)[l]}
  \put(13.05,36.87){\line(1,0){45.67}}
  \put(17.72,36.87){\vector(1,0){1.00}}
  \put(52.72,36.87){\vector(1,0){1.00}}
  \put(11.72,42.67){\makebox(0,0)[cc]{$p_1$}}
  \put(60.05,42.00){\makebox(0,0)[cc]{$p_1'$}}
  \put(11.72,7.67){\makebox(0,0)[cc]{$p_2$}}
  \put(30.72,29.67){\makebox(0,0)[cc]{$k_1$}}
  \put(55.72,27.67){\makebox(0,0)[cc]{$\pi^+$}}
  \put(60.29,7.19){\makebox(0,0)[cc]{$p_2'$}}
  \put(30.72,17.67){\makebox(0,0)[cc]{$k_2$}}
  \put(12.89,12.36){\line(1,0){45.67}}
  \put(12.89,12.03){\line(1,0){45.67}}
  \put(36.46,21.30){\oval(2.00,2.00)[r]}
  \put(36.49,23.69){\circle{2.34}}
  \put(38.25,24.00){\line(5,1){14.30}}
  \put(38.25,23.50){\line(5,-1){14.30}}
  \put(96.55,12.48){\oval(2.00,3.33)[lt]}
  \put(96.55,15.82){\oval(2.00,3.33)[rb]}
  \put(98.55,15.48){\oval(2.00,3.33)[lt]}
  \put(98.55,18.82){\oval(2.00,3.33)[rb]}
  \put(100.55,18.48){\oval(2.00,3.33)[lt]}
  \put(100.55,21.82){\oval(2.00,3.33)[rb]}
  \put(102.55,21.48){\oval(2.00,3.33)[lt]}
  \put(55.39,18.67){\makebox(0,0)[cc]{$\pi^-$}}
  \put(18.58,12.20){\vector(1,0){2.33}}
  \put(48.58,12.20){\vector(1,0){2.33}}
  \put(93.96,13.30){\oval(2.00,2.00)[r]}
  \put(93.96,15.30){\oval(2.00,2.00)[l]}
  \put(93.96,17.30){\oval(2.00,2.00)[r]}
  \put(93.96,19.30){\oval(2.00,2.00)[l]}
  \put(93.96,23.30){\oval(2.00,2.00)[l]}
  \put(93.96,25.40){\oval(2.00,2.00)[r]}
  \put(93.96,27.47){\oval(2.00,2.00)[l]}
  \put(93.96,29.47){\oval(2.00,2.00)[r]}
  \put(93.96,31.47){\oval(2.00,2.00)[l]}
  \put(93.96,33.47){\oval(2.00,2.00)[r]}
  \put(93.96,35.47){\oval(2.00,2.00)[l]}
  \put(75.39,36.47){\vector(1,0){1.00}}
  \put(110.39,36.47){\vector(1,0){1.00}}
  \put(89.06,23.00){\makebox(0,0)[cc]{$k_1$}}
  \put(102.73,16.34){\makebox(0,0)[cc]{$k$}}
  \put(70.72,36.50){\line(1,0){45.67}}
  \put(70.56,12.36){\line(1,0){45.67}}
  \put(70.56,12.03){\line(1,0){45.67}}
  \put(94.13,21.30){\oval(2.00,2.00)[r]}
  \put(76.25,12.20){\vector(1,0){2.33}}
  \put(106.25,12.20){\vector(1,0){2.33}}
  \put(94.14,12.15){\circle*{2.83}}
  \put(102.53,23.24){\line(3,5){5.00}}
  \put(102.53,23.24){\line(5,3){7.00}}
  \put(153.45,36.79){\oval(2.00,3.33)[lt]}
  \put(153.45,40.13){\oval(2.00,3.33)[rb]}
  \put(155.45,39.79){\oval(2.00,3.33)[lt]}
  \put(155.45,43.13){\oval(2.00,3.33)[rb]}
  \put(157.45,42.79){\oval(2.00,3.33)[lt]}
  \put(157.45,46.13){\oval(2.00,3.33)[rb]}
  \put(159.45,45.79){\oval(2.00,3.33)[lt]}
  \put(151.88,13.30){\oval(2.00,2.00)[r]}
  \put(151.88,15.30){\oval(2.00,2.00)[l]}
  \put(151.88,17.30){\oval(2.00,2.00)[r]}
  \put(151.88,19.30){\oval(2.00,2.00)[l]}
  \put(151.88,23.30){\oval(2.00,2.00)[l]}
  \put(151.88,25.40){\oval(2.00,2.00)[r]}
  \put(151.88,27.47){\oval(2.00,2.00)[l]}
  \put(151.88,29.47){\oval(2.00,2.00)[r]}
  \put(151.88,31.47){\oval(2.00,2.00)[l]}
  \put(151.88,33.47){\oval(2.00,2.00)[r]}
  \put(151.88,35.47){\oval(2.00,2.00)[l]}
  \put(133.31,36.47){\vector(1,0){1.00}}
  \put(168.31,36.47){\vector(1,0){1.00}}
  \put(128.64,36.50){\line(1,0){45.67}}
  \put(128.48,12.36){\line(1,0){45.67}}
  \put(128.48,12.03){\line(1,0){45.67}}
  \put(152.05,21.30){\oval(2.00,2.00)[r]}
  \put(134.17,12.20){\vector(1,0){2.33}}
  \put(164.17,12.20){\vector(1,0){2.33}}
  \put(151.91,36.90){\circle*{2.83}}
  \put(159.43,47.55){\line(3,5){5.00}}
  \put(159.43,47.55){\line(5,3){7.00}}
  \put(146.40,23.33){\makebox(0,0)[cc]{$k_2$}}
  \put(153.39,45.01){\makebox(0,0)[cc]{$k$}}
  \put(35.67,4.67){\makebox(0,0)[cc]{(a)}}
  \put(93.67,5.33){\makebox(0,0)[cc]{(b)}}
  \put(152.00,5.67){\makebox(0,0)[cc]{(c)}}
\end{picture}
\par
\caption{\small\emph{Diagrams for pion pair production in $e-\vec{p}$
  collisions.}}
\label{fig:diags}
\end{figure}
Let us now define the (virtual) exchange-photon 4-momenta in the problem:
\begin{displaymath}
  k_1 = p_1 - p'_1,
\qquad
  k_2 = p_2 - p'_2,
\end{displaymath}
together with
\begin{displaymath}
  k   = k_1 + k_2 = q_1 + q_2,
\qquad
  Q   = q_1 - q_2.
\end{displaymath}

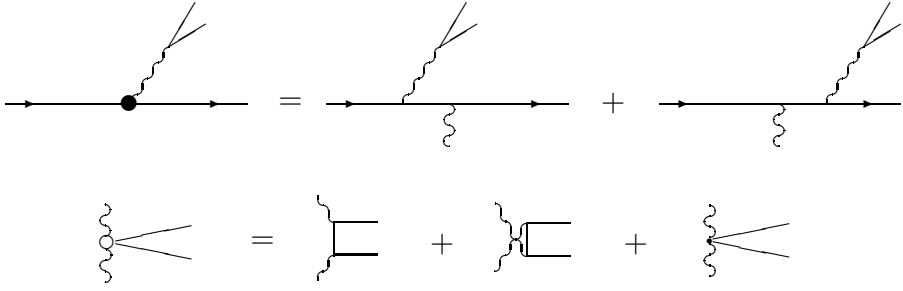
\begin{figure}
\unitlength=2.00pt
\special{em:linewidth 0.4pt}
\linethickness{0.4pt}
\begin{picture}(172.06,55.05)
  \put(21.61,5.27){\oval(2.00,2.00)[l]}
  \put(21.61,13.81){\oval(2.00,2.00)[l]}
  \put(21.61,15.81){\oval(2.00,2.00)[r]}
  \put(21.61,11.84){\oval(2.00,2.00)[r]}
  \put(21.78,7.27){\oval(2.00,2.00)[r]}
  \put(21.81,9.66){\circle{2.34}}
  \put(23.57,9.97){\line(5,1){14.30}}
  \put(23.57,9.47){\line(5,-1){14.30}}
  \put(27.62,35.96){\oval(2.00,3.33)[lt]}
  \put(27.62,39.30){\oval(2.00,3.33)[rb]}
  \put(29.62,38.96){\oval(2.00,3.33)[lt]}
  \put(29.62,42.30){\oval(2.00,3.33)[rb]}
  \put(31.62,41.96){\oval(2.00,3.33)[lt]}
  \put(31.62,45.30){\oval(2.00,3.33)[rb]}
  \put(33.62,44.96){\oval(2.00,3.33)[lt]}
  \put(2.81,35.84){\line(1,0){45.67}}
  \put(7.48,35.84){\vector(1,0){1.00}}
  \put(42.48,35.84){\vector(1,0){1.00}}
  \put(2.81,35.84){\line(1,0){45.67}}
  \put(26.08,36.07){\circle*{2.83}}
  \put(33.60,46.72){\line(3,5){5.00}}
  \put(33.60,46.72){\line(5,3){7.00}}
  \put(21.65,3.08){\oval(2.00,2.00)[r]}
  \put(51.34,9.67){\makebox(0,0)[cc]{=}}
  \put(56.67,36.00){\makebox(0,0)[cc]{=}}
  \put(78.93,35.96){\oval(2.00,3.33)[lt]}
  \put(78.93,39.30){\oval(2.00,3.33)[rb]}
  \put(80.93,38.96){\oval(2.00,3.33)[lt]}
  \put(80.93,42.30){\oval(2.00,3.33)[rb]}
  \put(82.93,41.96){\oval(2.00,3.33)[lt]}
  \put(82.93,45.30){\oval(2.00,3.33)[rb]}
  \put(84.93,44.96){\oval(2.00,3.33)[lt]}
  \put(63.47,35.84){\line(1,0){45.67}}
  \put(68.14,35.84){\vector(1,0){1.00}}
  \put(103.14,35.84){\vector(1,0){1.00}}
  \put(63.47,35.84){\line(1,0){45.67}}
  \put(84.91,46.72){\line(3,5){5.00}}
  \put(84.91,46.72){\line(5,3){7.00}}
  \put(117.67,36.00){\makebox(0,0)[cc]{+}}
  \put(159.07,35.96){\oval(2.00,3.33)[lt]}
  \put(159.07,39.30){\oval(2.00,3.33)[rb]}
  \put(161.07,38.96){\oval(2.00,3.33)[lt]}
  \put(161.07,42.30){\oval(2.00,3.33)[rb]}
  \put(163.07,41.96){\oval(2.00,3.33)[lt]}
  \put(163.07,45.30){\oval(2.00,3.33)[rb]}
  \put(165.07,44.96){\oval(2.00,3.33)[lt]}
  \put(126.39,35.84){\line(1,0){45.67}}
  \put(131.06,35.84){\vector(1,0){1.00}}
  \put(166.06,35.84){\vector(1,0){1.00}}
  \put(126.39,35.84){\line(1,0){45.67}}
  \put(165.05,46.72){\line(3,5){5.00}}
  \put(165.05,46.72){\line(5,3){7.00}}
  \put(149.06,28.83){\oval(2.00,2.00)[l]}
  \put(149.06,30.83){\oval(2.00,2.00)[r]}
  \put(149.06,32.85){\oval(2.00,2.00)[l]}
  \put(149.06,34.85){\oval(2.00,2.00)[r]}
  \put(86.73,32.84){\oval(2.00,2.00)[l]}
  \put(86.73,34.84){\oval(2.00,2.00)[r]}
  \put(86.73,30.87){\oval(2.00,2.00)[r]}
  \put(86.62,28.82){\oval(2.00,2.00)[l]}
  \put(64.83,13.50){\line(0,-1){6.17}}
  \put(64.96,13.46){\line(1,0){8.17}}
  \put(64.83,7.33){\line(1,0){8.17}}
  \put(101.31,13.17){\line(0,-1){6.17}}
  \put(101.31,13.26){\line(1,0){8.17}}
  \put(101.31,7.00){\line(1,0){8.17}}
  \put(135.94,15.62){\oval(2.00,2.00)[r]}
  \put(135.94,13.62){\oval(2.00,2.00)[l]}
  \put(135.94,11.62){\oval(2.00,2.00)[r]}
  \put(135.94,4.23){\oval(2.00,2.00)[r]}
  \put(135.94,6.23){\oval(2.00,2.00)[l]}
  \put(135.94,8.33){\oval(2.00,2.00)[r]}
  \put(136.58,10.25){\line(5,1){14.30}}
  \put(136.58,9.75){\line(5,-1){14.30}}
  \put(85.33,9.33){\makebox(0,0)[cc]{+}}
  \put(122.00,9.33){\makebox(0,0)[cc]{+}}
  \put(95.26,5.87){\oval(2.00,3.33)[rb]}
  \put(97.26,5.53){\oval(2.00,3.33)[lt]}
  \put(97.26,8.87){\oval(2.00,3.33)[rb]}
  \put(99.26,8.53){\oval(2.00,3.33)[lt]}
  \put(99.26,11.87){\oval(2.00,3.33)[rb]}
  \put(101.26,11.53){\oval(2.00,3.33)[lt]}
  \put(62.82,2.44){\oval(2.00,3.33)[lt]}
  \put(62.82,5.78){\oval(2.00,3.33)[rb]}
  \put(64.82,5.44){\oval(2.00,3.33)[lt]}
  \put(62.87,15.03){\oval(2.00,3.33)[rt]}
  \put(64.87,15.10){\oval(2.00,3.33)[lb]}
  \put(62.87,18.36){\oval(2.00,3.33)[lb]}
  \put(99.31,8.59){\oval(2.00,3.33)[rt]}
  \put(101.31,8.66){\oval(2.00,3.33)[lb]}
  \put(99.31,11.92){\oval(2.00,3.33)[lb]}
  \put(97.36,11.91){\oval(2.00,3.33)[rt]}
  \put(97.36,15.29){\oval(2.00,3.33)[lb]}
  \put(95.41,15.32){\oval(2.00,3.33)[rt]}
  \put(135.93,9.97){\circle*{1.09}}
\end{picture}
\par
\caption{\small\emph{Decoding of the notation used in
  Fig.~\protect\ref{fig:diags}.}}
\label{fig:decode}
\end{figure}

Exploiting the infinite-momentum frame approach, we may neglect the
contribution from the diagram of Fig.~\ref{fig:diags}c and thus write down the
matrix element in the following form
\begin{equation}
  \mathcal{M}_\mathrm{jet-2}
  =
  \mathcal{M}_D + \mathcal{M}_B
  =
  \frac{(4\pi\alpha)^2}{k_1^2} \;
  \bar{u}(p_1') \gamma_\mu u(p_1) \;
  \bar{u}(p_2') J^\mu u(p_2).
\label{eq:4}
\end{equation}
{}From the double-photon (D) and bremsstrahlung (B) mechanisms for pion pair
production, the current receives contributions of the form:
\begin{eqnarray}
  J^\mu   = J^\mu_D + J^\mu_B,
\quad
  J^\mu_D = \frac{1}{k_2^2} \; \mathcal{M}^{\mu\nu} \gamma_\nu,
\quad
  J^\mu_B = \frac{F(k^2)}{k^2} \; \mathcal{O}^{\mu\nu} Q_\nu,
\end{eqnarray}
where the pion form factor is taken to be of the form
\begin{displaymath}
  F(k^2)
  =
  \frac{m_\rho^2}
       {k^2-m_\rho^2+\im m_\rho\Gamma_\rho}.
\end{displaymath}
We have introduced here the following tensors (see Fig.~\ref{fig:decode} for
the graphical correspondence):
\begin{eqnarray}
  \mathcal{M}^{\mu\nu}
  &=&
  \frac{(2q_1-k_1)^\mu (k_2-2q_2)^\nu}{\chi_1}
  +
  \frac{(2q_1-k_2)^\nu (k_1-2q_2)^\mu}{\chi_2}
  -
  2 g^{\mu\nu}
\label{eq:6}
\\[1ex]
\noalign{\noindent (defining $\chi_1=k_1^2-2k_1{\cdot}q_1$ and
$\chi_2=k_1^2-2k_1{\cdot}q_2$) and\rule[-2ex]{0pt}{0pt}}
  \mathcal{O}^{\mu\nu}
  &=&
  \gamma^\nu \, \frac{\hat{p}_2+\hat{k}_1+M}{(p_2+k_1)^2-M^2} \, \gamma^\mu
  +
  \gamma^\mu \, \frac{\hat{p}_2'-\hat{k}_1+M}{(p_2'-k_1)^2-M^2} \, \gamma^\nu,
\end{eqnarray}
which are subject to the gauge conditions
\begin{displaymath}
  \mathcal{M}^{\mu\nu} k_{1\mu}
  =
  0
  =
  \mathcal{M}^{\mu\nu} k_{2\nu},
\end{displaymath}
\begin{equation}
  \bar{u}(p_2') \mathcal{O}^{\mu\nu} u(p_2) \, k_{1\mu}
  =
  0
  =
  \bar{u}(p_2') \mathcal{O}^{\mu\nu} u(p_2) \, k_\nu.
\end{equation}

We introduce the standard Sudakov parametrisation of the 4-momenta in the
problem
\begin{equation}
\arraycolsep0.2em
\begin{array}{lcccccc}
  q_i  &=& x_i\tilde{p}_2    &+& \beta_i\tilde{p}_1  &+& q_{i\bot},
\\
  k_1  &=& \alpha\tilde{p}_2 &+& \beta\tilde{p}_1    &+& k_{1\bot},
\\
  p_2' &=& x\tilde{p}_2      &+& \beta_2'\tilde{p}_1 &+& p_{2\bot}',
\end{array}
\end{equation}
where we have used the following:
\begin{displaymath}
  \tilde{p}_2 = p_2 - \frac{M^2}{s} p_1,
  \quad
  \tilde{p}_1 = p_1 - \frac{m_e^2}{s} p_2,
  \quad
  s           = 2 p_1{\cdot}p_2 \gg M^2.
\end{displaymath}
Here
\begin{displaymath}
  x_i = \frac{q_{iz}+E_i}{2E}
\qquad \mathrm{and} \qquad
  x = \frac{p'_{2z}+E'_2}{2E}
\end{displaymath}
are the pion and scattered-proton energy fractions ($x+x_1+x_2=1$),
$q_{i\bot}$ and $p_{2\bot}'$ are the 4-momenta transverse to the beam axes.
The corresponding euclidean 2-vectors are $\vecc{q}_i$ and $\vecc{p}_2'$.

The small parameters, $\beta$, may be expressed via these vectors:
\begin{equation}
  \beta_i=\frac{\vecc{q}_i^2+m^2}{sx_i},
\qquad
  \beta_2'=\frac{\vecc{p}_2^{\prime2}+M^2}{sx}.
\end{equation}
In terms of these variables, the cross-section and the phase volume is
\begin{displaymath}
  \d\sigma
  =
  \frac{1}{8s} \sum |\mathcal{M}|^2 \, \d\Gamma,
\end{displaymath}
\begin{eqnarray}
  \d\Gamma
  &=&
  \frac{\d^3p_1' \, \d^3p_2' \, \d^3q_1 \, \d^3q_2}
       {(2\pi)^8 2E_1' 2E_2' 2E_1 2E_2}
  \;
  \delta^4(p_1 + p_2 - p_1' - p_2' - q_1 - q_2)
\nonumber \\
  &=&
  \frac{\d^2\vecc{q}_1 \, \d^2\vecc{q}_2 \, \d^2\vecc{k}_1 \, \d{x}_1 \, \d{x}_2}
       {(2\pi)^8 8s x x_1 x_2}.
\end{eqnarray}

The jet-2 kinematics permits us to use the Gribov representation for the
exchange-photon Green function, see Eq.~(\ref{eq:4}):
\begin{equation}
  \frac{g^{\mu\nu}}{k_1^2}
  =
  \frac{1}{k_1^2}
  \left[
  g^{\mu\nu}_\bot + \frac{2}{s} (p_1^\mu p_2^\nu + p_1^\nu p_2^\mu)
  \right]
  \simeq
  \frac{2}{s k_1^2}p_1^\mu p_2^\nu,
\end{equation}
and from the gauge condition,
$J^{\mu}k_{1\mu}=J^\mu({\beta}p_1+k_{1\bot})_\mu=0$, it follows that
\begin{equation}
  J^\mu p_{1\mu}
  =
  -\frac{s}{s_2} J^\mu k_{1\bot\mu}.
\end{equation}
Here we have defined $\tilde{s}_2=(p_2'+q_1+q_2)^2=s_2+M^2$ and denoted the
invariant jet-mass squared by
\begin{eqnarray}
  s_2 = s\beta =
  \frac{\vecc{p}_1^{\prime2} + (1-x)M^2}{x}
  + \frac{\vecc{q}_1^2 + m^2}{x_1}
  + \frac{\vecc{q}_2^2 + m^2}{x_2}.
\end{eqnarray}

For the modulus squared of the matrix element summed over spin states and
averaged over the azimuthal angle of the virtual photon, we obtain
\begin{equation}
  \sum |\mathcal{M}|^2
  =
  -\frac{(4\pi\alpha)^4}{(k_1^2)^2}
  \frac{4s^2\vecc{k}_1^2}{s_2^2}
  \;
  \Tr[(\hat{p}_2'+M)
  J^\mu_\bot
  (\hat{p}_2+M)(1-\gamma_5\hat{a})
  \tilde{J}_{\mu\bot}].
\end{equation}
This expression contains both the charge-even and charge-odd contributions,
in addition to which it contains the spin correlation term associated with
the proton polarisation vector, $a$. The cross-section acquires the
well-known \WW enhancement factor:
\begin{equation}
  \int \frac{\d^2 \vecc{k}_1 \vecc{k}_1^2}{\pi(k_1^2)^2}
  =
  L,
\end{equation}
where the quantity $L$ stands for a \emph{large} logarithm, whose value
depends on the type of the initial particle with momentum $p_1$. In the case where it is a proton we have
\begin{eqnarray*}
  k_1^2
  &=&
  -\left[ \vecc{k}_1^2 + \left( M \frac{s_1}{s} \right)^2 \right],
  \qquad
  L = L_p = \ln\left( \frac{ms}{Ms_1} \right)^2,
\end{eqnarray*}
\begin{eqnarray}
  s_1
  &=&
    \frac{\vecc{p}_1^{\prime2}}{x}
  + \frac{\vecc{q}_1^2+m^2}{x_1}
  + \frac{\vecc{q}_2^2+m^2}{x_2},
  \quad
  M = M_p,
  \quad
  m = m_\pi.
\end{eqnarray}
For the case where it is an electron we have
\begin{equation}
  k_1^2 = - \left[ \vecc{k}_1^2 + \left( m_e\frac{s_2}{s} \right)^2 \right],
  \qquad
  L = L_e = \ln\left( \frac{ms}{m_es_2} \right)^2.
\end{equation}
The main (logarithmic) contribution to the cross-section comes from the
region $|\vecc{k}_1|\ll|\vecc{q}_i|$, which provides the relation between the
transverse components of jet particles: $\vecc{q}_1+\vecc{q}_2+\vecc{p}_2'=0$.

Upon doing a standard matrix element calculus spin-dependent part of the
differential cross-section
for the jet-2 kinematics is found to be
\begin{eqnarray}
  \d\sigma^{\mathrm{spin}}_\mathrm{jet-2}
  &=&
  \frac{\alpha^4L_e}{\pi^3s_2^3}
  \frac{\Im F^*(k^2)M}{k_2^2 k^2}
\nonumber\\
  &&\hspace*{-1em} \mbox{} \times
  \left[
    A(\vecc{q}_1,x_1;\vecc{q}_2,x_2) (\vecc{q}_1\vprod\vecc{a})_z
    -
    A(\vecc{q}_2,x_2;\vecc{q}_1,x_1) (\vecc{q}_2\vprod\vecc{a})_z
  \right]
\nonumber\\
  &&\hspace*{-1em} \mbox{} \times
  \frac{\d^2\vecc{q}_1 \, \d^2\vecc{q}_2 \, \d{x}_1 \, \d{x}_2}
       {x_1 x_2 x},
\end{eqnarray}
with the expression for $A$ given in the Appendix.
Its reduced version for a back-to-back kinematics is quoted below.

\section{The general form of $\gamma\gamma\to\pi^+\pi^-$ amplitude}

The tensor describing the conversion of two photons into a pion pair can be
presented in a form that may be interpreted as an expansion in pion
momenta\footnote{Here $a_0$ and $a_2$ may be associated with the partial-wave
decomposition of pion--pion scattering amplitudes (not to be confused with
the contribution of states with isotopic spin 0 and 2). The quantity $a_0$ at
threshold is connected with the pion polarisability
$\alpha_\pi=a_0/(8{\pi}m)$.}:
\begin{equation}
  \mathcal{M}^{\mu\nu}
  =
  a_0 \mathcal{L}_0^{\mu\nu} + a_2 \mathcal{L}_2^{\mu\nu},
\end{equation}
where
\begin{displaymath}
  a_0 = \frac{\chi_1+\chi_2}{\chi_1\chi_2},
\quad
  a_2 = \frac{2}{\chi_1\chi_2},
\end{displaymath}
\begin{eqnarray*}
  \mathcal{L}_0^{\mu\nu}
  &=&
  k_1{\cdot}k_2g^{\mu\nu}-k_1^\nu k_2^\mu,
\\
  \mathcal{L}_2^{\mu\nu}
  &=&
  - k_1{\cdot}k_2 Q^\mu Q^\nu - Q{\cdot}k_1 (Q^\mu k_1^\nu + Q^\nu k_1^\mu)
\\
  && \hspace{1em}\mbox{}
  + Q{\cdot}k_1 Q^\nu (k_1 + k_2)^\mu + (Q{\cdot}k_1)^2 g^{\mu\nu},
\end{eqnarray*}
and
\begin{displaymath}
  \mathcal{L}_i^{\mu\nu} k_{1\mu}
    = 0 =
  \mathcal{L}_i^{\mu\nu} k_{2\nu}.
\end{displaymath}

Taking into account final-state interactions, one expects these amplitudes,
as well as the amplitude for the conversion of a single photon into a pion
pair, to acquire the following phases:
\begin{equation}
  a_0    \to a_0    \e^{\im\delta_0},\quad
  F(k^2) \to F(k^2) \e^{\im\delta_1},\quad
  a_2    \to a_2    \e^{\im\delta_2}.
\end{equation}
The phases $\delta_{0,1,2}$ are associated with states having orbital angular
momentum equal to $0,1,2$ respectively. It is useful to note that the third
possible gauge-invariant structure,
\begin{equation}
  \mathcal{L}_3^{\mu\nu}
  =
    Q{\cdot}k_1 (k_1^2 g^{\mu\nu} - k_1^\mu k_1^\nu)
  + Q^\nu (k_1^2 k_2^\mu - k_1{\cdot}k_2 k_1^\mu),
\end{equation}
which may be thought of as a pair in a state with unit orbital angular
momentum, is not realised in the double-photon
channel\footnote{We are grateful
to V.~Serbo and I.~Ginzburg for discussion on this point.}.

\section{Jet-1 kinematics}

Similar considerations may be applied to the jet-1 kinematics: where a jet
moving close to the initial-electron direction consists of a scattered
electron and a pion pair. Here we use the following parametrisation of the
momenta:
\begin{equation}
\arraycolsep0.2em
\begin{array}{lcccccc}
  q_i  &=& \alpha_i\tilde{p}_2  &+& x_i\tilde{p}_1   &+& q_{i\bot},
\\
  p_1' &=& \alpha_1'\tilde{p}_2 &+& x\tilde{p}_1     &+& p_{1\bot}',
\\
  k_1  &=& \alpha\tilde{p}_2    &+& \beta\tilde{p}_1 &+& k_{1\bot},
\nonumber
\end{array}
\end{equation}
with
\begin{equation}
  \alpha_i  = \frac{\vecc{q}_i^2+m^2}{sx_i}
\qquad\mathrm{and}\qquad
  \alpha_1' = \frac{\vecc{p}_1^{\prime2}}{sx},
\end{equation}
where $x$ and $x_i$ are the energy fractions of a scattered electron and
produced pions ($x+x_1+x_2=1$); $\vecc{p}_1'$ and $\vecc{q}_i$ are their
transverse momenta, obeying $\vecc{p}_1'+\vecc{q}_1+\vecc{q}_2=0$.

We thus obtain
\begin{eqnarray}
  \sum|\mathcal{M}_D|^2
  &=&
  -\frac{16(4\pi\alpha)^4\vecc{k}_1^2s^2}
        {(k_1^2k_2^2s_1)^2} \;
  \frac14
  \Tr\left[
       \hat{p}_1' \hat{R}^\sigma_\bot \hat{p}_1 \hat{\tilde R}_{\sigma\bot}
     \right],
\nonumber\\
  \sum|\mathcal{M}_B|^2
  &=&
  -\frac{16(4\pi\alpha)^4\vecc{k}_1^2|F(k^2)|^2s^2}
        {(k_1^2k^2s_1)^2} \;
  \frac14
  \Tr\left[
       \hat{p}_1' \hat B^\sigma_\bot \hat{p}_1 \hat{\tilde{B}}_{\sigma\bot}
     \right],
\\
  2\Re \sum(\mathcal{M}_B \mathcal{M}_D^*)
  &=&
  -\Re \frac{32(4\pi\alpha)^4\vecc{k}_1^2F(k^2) \e^{\im\delta_1}}
            {(k_1^2)^2k_2^2k^2} \;
  \frac14
  \Tr\left[
       \hat{p}_1' \hat B^\sigma_\bot \hat{p}_1 \hat{\tilde R}_{\sigma\bot}
     \right],
\nonumber
\end{eqnarray}
where
\begin{displaymath}
  s_1
  =
  s \alpha
  =
    \frac{\vecc{p}_1^{\prime2}}{x}
  + \frac{\vecc{q}_1^2+m^2}{x_1}
  + \frac{\vecc{q}_2^2+m^2}{x_2}
\end{displaymath}
is the jet invariant-mass squared,
\begin{displaymath}
  k = q_1 + q_2,
\qquad
  k_1^2 = - \left[ \vecc{k}_1^2 + M^2\left(\frac{s_1}{s}\right)^2 \right],
\end{displaymath}
and
\begin{equation}
  \hat B_\sigma
  =
  \frac{1}{s}
    \left(
      \hat{Q}\hat{p}_2\gamma_\sigma
    + \frac{1}{x} \gamma_\sigma \hat{p}_2 \hat{Q}
    \right)
  + \frac{2}{s_1x} p_{1\sigma}' \hat{Q},
\end{equation}
and $\hat{R}_\sigma=\mathcal{M}_{\sigma\mu}\gamma^\mu$, see Eq.~(\ref{eq:6}).
Proton spin correlations are absent here since in the \WW approximation the
exchange photon cannot carry spin information.

Taking into account the strong phases in the pion amplitudes, the expression
for differential cross-section is modified as follows:
\begin{eqnarray}
  \d\sigma^\mathrm{odd}_\mathrm{jet-1}
  &=&
  - \frac{\alpha^4L_p}{\pi^3s_1^2k_2^2k^2}
  \left\{
    I \, \Re \left[ F(k^2) e^{\im(\delta_1-\delta_2)} \right]
  \right.
\label{eq:25}
\nonumber\\
  && \hspace*{-2em}
  \left.
  \null
         +
  \tilde I \, \Re \left[ F(k^2)
    \left(
      e^{\im(\delta_1-\delta_0)}
    - e^{\im(\delta_1-\delta_2)}
    \right) \right]
  \right\}
  \frac{\d^2 \vecc{q}_1 \, \d^2 \vecc{q}_2 \, \d{x}_1 \, \d{x}_2}
       {x x_1 x_2}.
\end{eqnarray}

\section{Back-to-back kinematics}

For the kinematics with $|\vecc{p}'_1|\to0$ the charge-odd part of the
cross-section takes the form:
\begin{eqnarray}
  \left.
  \d\sigma^\mathrm{odd}_\mathrm{jet-1}
  \right|^{(\pi\pi)}_{\vecc{q}_2=-\vecc{q}_1}
  &=&
  \frac{\alpha^4L_p}{\pi^3}
  \frac{\vecc{q}_1^2-\vecc{q}_2^2}{(\vecc{q}_1+\vecc{q}_2)^2}
  \frac{(x_1x_2)^2}{(\vecc{q}_1^2+m^2)^3 x (1-x)^5}
\\
  &\times&
  \Biggl\{
    -2x\Re\left[
            F(k^2) \left(
                     \e^{\im(\delta_1-\delta_0)}
                   - \e^{\im(\delta_1-\delta_2)}
                   \right)
          \right]
\nonumber\\
  && \mbox{}
  + \Re\left[ F(k^2)\e^{\im(\delta_1-\delta_2)} \right]
      \left( 1 + x^2
         - \frac{m^2}{\vecc{q}_1^2+m^2}(1+x)^2
      \right)
  \Biggr\}
\nonumber\\
  &\times&
  \d^2\vecc{q}_1 \, \d^2\vecc{q}_2 \, \d{x}_1 \, \d{x}_2,
\nonumber
\end{eqnarray}
for $\e^{-\im\delta_i}=1$, this is in agreement with the expression obtained
in~\cite{Serbo:1974x1, Budnev:1975x1}. We also include here a similar
expression for muon-pair production in this limit:
\begin{eqnarray}
  \d\sigma^\mathrm{odd}_\mathrm{jet-1}|^{(\mu\mu)}_{\vecc{q}_2=-\vecc{q}_1}
  &=&
  \frac{\alpha^4L_p}{\pi^3}
  \frac{\vecc{q}_1^2-\vecc{q}_2^2}
       {(\vecc{q}_1+\vecc{q}_2)^2}
  \frac{x_1 x_2}
       {(\vecc{q}_1^2+m^2)^4 x (1-x)^5}
\nonumber\\
  &&\hspace*{-1em} \mbox{} \times
  \left\{
    \left[ (1-x)^2(\vecc{q}_1^2+m^2)-2\vecc{q}_1^2x_1x_2 \right](1+x^2)
    + 4xx_1x_2m^2
  \right\}
\nonumber\\
  &&\hspace*{-1em} \mbox{} \times
  \d^2\vecc{q}_1 \, \d^2\vecc{q}_2 \, \d{x}_1 \, \d{x}_2.
\end{eqnarray}

For the kinematics where
$\vecc{p}^{\prime2}_1=(\vecc{q}_1+\vecc{q}_2)^2\ll\vecc{q}^2_1$ we have (using
the $\delta$-function approximation for $\Im{F}^*$):
\begin{eqnarray}
  \d\sigma^\mathrm{spin}_\mathrm{jet-2}|^{(\pi\pi)}_{\vecc{q}_2=-\vecc{q}_1}
  &=&
  \frac{2\alpha^4L_e}{\pi^2}
  M(\vecc{q}_1\vprod\vecc{a})_z
\nonumber\\
  && \mbox{} \times
  \frac{(xx_1x_2)^2}
       {v(M^2(1-x)^2+\vecc{p}_2^{\prime 2})(1-x)^3(vx+M^2x_1x_2)^3}
\nonumber\\
  && \mbox{} \times
  \left[
    v - m^2(1+x)
    - \frac{M^2x_1x_2}{x}
    - \frac{v}{4} \frac{(1-x)^3}{x_1x_2}
  \right]
\\
  && \mbox{} \times
  \delta \left( 1 - \frac{v(1-x)^2}{x_1x_2m_\rho^2} \right)
  \;
  \d^2\vecc{q}_1 \, \d^2\vecc{q}_2 \, \d{x}_1 \, \d{x}_2,
\nonumber
\end{eqnarray}
where $v=\vecc{q}_1^2+m^2$.

Let us now discuss the accuracy of the formul{\ae} presented, determined, of
course, by the omitted terms: they are of order
\begin{equation}
  \frac{m^2}{s}, \quad
  \frac{s_i}{s}, \quad \mbox{and}\quad
  \frac{1}{L},
\end{equation}
as compared to unity. For the DESY experimental conditions the accuracy is
better than $10\%$. For the inclusive set-up, one must consider the 3-pion
production process. Compared with 2-pion production, this has a phase-volume
suppression factor
\begin{displaymath}
  \int \frac{\d^2\vecc{q}}{{\pi}M^2}
  \sim
  \left( \frac{m}{M} \right)^2
  \sim
  10^{-2}.
\end{displaymath}

For intermediate energies (such as at VEPP-2M and DA$\Phi$NE), despite the
rather small suppression factor $m^2/s\sim0.01$ for the theoretical
background of the remaining Feynman diagrams (there are six gauge-invariant
sets, of which we have considered just two), the \WW enhancement factor (a
large logarithm) and the specific choice of kinematics
($|\vecc{q}_1+\vecc{q}_2|\sim0$) provides an accuracy of the same order for
jet-1 kinematics as obtained for jet-2 type.

\section{Conclusion}

Using conditions of the back-to-back kinematics as well as the $\rho$ meson
dominance approximation
the charge-odd contribution to the spectrum in a jet-1 kinematics for the
case $|\vecc{p}_1'|\ll|\vecc{q}_1|$ may be cast into the following form:
\begin{eqnarray}
  m_\rho^2\frac{\d\sigma^{\mathrm{odd}(\pi^+\pi^-)}_\mathrm{jet-1}}
  {\d^2\vecc{q}_1 \, \d{x}_1 \, \d{x}_2}
  &=&
  0.7 \cdot 10^{-2} \, \mathrm{nb}
\label{eq:main}
\\
  && \hspace{1ex} \mbox{} \times
  \frac{\vecc{q}_1^2-\vecc{q}_2^2}{(\vecc{q}_1+\vecc{q}_2)^2} \;
  [f_1\sin(\delta_0-\delta_1) + f_2 \sin(\delta_2-\delta_1)],
\nonumber
\end{eqnarray}
with $f_1$ and $f_2$ being smooth functions of order unity:
\begin{equation}
  f_1 = -\frac{2}{x_1+x_2},
\qquad
  f_2 = \frac{(1+x)^2}{x(x_1+x_2)}.
\label{ff}
\end{equation}
The contribution to the spin-dependent part in the same kinematics can be
cast into the form
\begin{displaymath}
  \frac{\d\sigma^{\mathrm{spin}(\pi^+\pi^-)}_\mathrm{jet-2}}
       {\d\phi \, \d{x}_1 \, \d{x}_2}
  =
  1.8 \cdot 10^{-2}\,\mathrm{nb} \,
  \cdot |a|f_3 \frac{\d\vecc{p}_1^{\prime2}}{M^2(1-x)^2 +\vecc{p}_1^{\prime2}}
  \cdot \sin\phi,
\end{displaymath}
with
\begin{equation}
  f_3
  =
  \frac{1.22x}{\sqrt{x_1x_2}}
  \frac{[ x_1x_2 x - 1.49(x_1+x_2)^2x_1x_2 - 0.25 x(x_1+x_2)^3]}
       {[ x + 1.49(x_1+x_2)^2 ]^3 }.
\label{eq:mainspin}
\end{equation}
Here $|a|$ is the degree of target transverse polarisation; $\phi$ is the
azimuthal angle between the directions of target polarisation and transverse
momentum, $\vecc{q}_1$, of a negative pion. Typical values of the functions
$f_i$ for a set of $x_1,x_2$ values are given in Table~\ref{tab:num} for
HERMES conditions ($s=60\,$GeV$^2$ and
$|\vecc{q}_1|\approx|\vecc{q}_2|\sim0.5\div1.5\,$GeV).

The calculation presented above has been carried out within the framework of
QED. In the case of jet-1 kinematics it is possible to replace photon exchange
between the pion and nucleon by that of a \emph{pomeron} in the double-photon
amplitudes. This results in an enhancement factor $\alpha_s/\alpha$:
$\sigma_0\to(\alpha_s/\alpha)\sigma_0$.

A visible effect of the asymmetry in pion pair production was measured at
Cornell\footnote{We thank V.G.~Serbo for bringing this to our attention.}
where the number of most
energetic $\pi^-$ moving along the $e^-$ direction exceeds that of most
energetic $\pi^+$ in the same direction.

The formul{\ae} given above may be applied to pion pair production at $e^+e^-$
colliders. Besides the effect of contributions of the annihilation type,
Feynman diagrams fall with the growth of the total centre-of-mass energy
($\sqrt{s}$) since they are of order $m^2/s<3\%$ for $J/\psi$ and $B$
factories. Nevertheless, charge-odd effects in $\pi^+\pi^-$ production may
definitely be measured at $\Phi$-factories by taking advantage of the
kinematics discussed above. Moreover, charge-odd effects in $K_S$ and $K_L$
pair production may clearly be seen at $J/\psi$ and $B$ factories.

\begin{table}
\newcommand{\M}{\hphantom{-}}
\newcommand{\Z}{\hphantom{0}}
\begin{displaymath}
\begin{array}{|c|c||c|c|c|c|}
\hline
  x_1  &  x_2  &   f_1   &   f_2   &   f_3   &   f_4  \\ \hline
  0.2  &  0.2  & -5.000  & 10.667  &\M0.030  &\M0.000 \\ \hline
  0.2  &  0.4  & -3.333  &\Z8.167  & -0.068  & -0.173 \\ \hline
  0.2  &  0.6  & -2.500  &\Z9.000  & -0.053  & -0.141 \\ \hline
  0.3  &  0.4  & -2.857  &\Z8.048  & -0.075  & -0.065 \\ \hline
\end{array}
\end{displaymath}
\caption{\small\emph{The values of the functions} $f_1$, $f_2$
[Eq.~(\protect\ref{ff})],
$f_3$ [Eq.~(\protect\ref{eq:mainspin})] \emph{and}
$f_4$ [Eq.~(\protect\ref{f4})]
  \emph{for typical HERMES conditions} ($|\vecc{p}_1'|\ll|\vecc{q}_1|$,
  $0.2\,\mathrm{GeV}<|\vecc{q}_1|<1.2\,\mathrm{GeV}$, $0.2<x_i<0.6$).}
\label{tab:num}
\end{table}

\appendix
\section*{Appendix}

The general expression in a jet-1 kinematics for the charge-odd contribution
to a pion pair production cross-section in the \WW approximation is given in
Eq.~(\ref{eq:25}) with the quantities $I$ and $\tilde{I}$:
\begin{eqnarray}
  I &=& - \vecc{Q}\vecc{p}'_1 + \frac{(1+x)(x_1-x_2)}{x} p_1{\cdot}p'_1
  + \frac{1+x}{2x_1x_2} \vecc{b}\vecc{Q}
  - \frac{x_1-x_2}{2x_1x_2} \vecc{b}\vecc{p}'_1
\nonumber\\
  &&  - \frac{\vecc{p}'_1\vecc{a}}{x_1x_2x}
    \frac{(2p_1{\cdot}Q 2 p'_1{\cdot}Q - Q^2 2p_1{\cdot}p'_1)}
         {s_1^2}
  + \frac{2(-\vecc{p}_1^{\prime2} Q{\cdot}p_1 + \vecc{Q}\vecc{p}'_1 p_1{\cdot}p'_1)}
         {x s_1}
\nonumber\\
  && - \frac{\vecc{Q}\vecc{a}}{x_1x_2s_1} (p_1{\cdot}Q+p'_1{\cdot}Q)
  + \frac{\vecc{p}'_1\vecc{b}}{x_1x_2s_1} (x Q{\cdot}p_1+Q{\cdot}p'_1-(x_1-x_2)p_1{\cdot}p'_1)
\nonumber\\
  && - \frac{(1-x)Q^2}{2s_1} \frac{\vecc{p}'_1\vecc{a}}{x_1x_2x}
  + \frac{x_1-x_2}{xx_1x_2s_1} Q{\cdot}p_1 \vecc{p}'_1\vecc{a},
\end{eqnarray}
where
\begin{displaymath}
  \vecc{b} = x_2 \vecc{q_1} + x_1 \vecc{q_2},
\qquad
  \vecc{a} = x_2 \vecc{q_1} - x_1 \vecc{q_2},
\end{displaymath}
\begin{displaymath}
  k_2^2 = -2p_1{\cdot}p'_1 = -\frac{\vecc{p}_1^{2\prime}}{x},
\qquad
  2p'_1{\cdot}Q = x 2p_1{\cdot}Q-2 \vecc{Q}\vecc{p}'_1,
\end{displaymath}
\begin{displaymath}
  2p_1{\cdot}Q = \frac1{x_1} (\vecc{q}_1^2+m^2) - \frac1{x_2} (\vecc{q}_2^2+m^2)
\end{displaymath}
\begin{equation}
  Q^2 = 4m^2 - k^2,
\qquad
  k^2 = \frac1{x_1x_2} \left( m^2(1-x)^2 + \vecc{a}^2 \right),
\end{equation}
and
\begin{eqnarray}
  \tilde I
  &=&
  -\frac{2}{1-x} \vecc{p}'_1 \vecc{Q}
  + \vecc{p}_1^{\prime2}
    \left[
      \frac{(1-x+x^2)(x_1-x_2)}{x^2(1-x)} - \frac{4p_1{\cdot}Q}{s_1x(1-x)}
    \right]
\nonumber\\
  && \mbox{}
  + \frac{1+x}{x^2(1-x)} \frac{\vecc{p}_1^{\prime2}\,
  \vecc{p}'_1 \vecc{Q}}{s_1}
  + \frac{x_1-x_2}{x^2(1-x)} \frac{(\vecc{p}_1^{\prime2})^2}{s_1}.
\end{eqnarray}

The quantity $A$, relevant for the spin asymmetry, is
\begin{eqnarray}
  \mathcal{A}(\vecc{q}_1,x_1;\vecc{q}_2,x_2)
  &=&
  - s_2 \frac{1-x}{x}
  + \frac{(1-x)^2Q^2}{2xx_1}
  + \frac{x_1-x_2}{2xx_1} (-2xp_2'{\cdot}Q+2p_2{\cdot}Q)
\nonumber\\
  && \hspace*{-6em} \mbox{}
  + \frac{2}{xx_1}
    [
      x_2\vecc{q}_1^2 + x_1\vecc{q}_2^2 - (1-x)\vecc{q}_1\vecc{q}_2
    ]
  - \frac{2x_2}{xx_1} (\vecc{q}_1^2 - \vecc{q}_2^2)
\nonumber\\
  && \hspace*{-6em} \mbox{}
  + \left[
      \frac{x_1}{x_2} (\vecc{q}_2^2+m^2) - \frac{x_2}{x_1} (\vecc{q}_1^2+m^2)
    \right]
  \left[
     \frac{x_1-x_2}{xx_1} + \frac{2}{s_2xx_1}(\vecc{q}_2^2 - \vecc{q}_1^2)
  \right].
\end{eqnarray}


\end{document}